# Modeling and Simulations of a Single-Spin Measurement using MRFM

G.P. Berman, F. Borgonovi, V.N. Gorshkov, V.I. Tsifrinovich

*Abstract*— We review the quantum theory of a single spin magnetic resonance force microscopy (MRFM). We concentrate on the novel technique called oscillating cantilever-driven adiabatic reversals (OSCAR), which has been used for a single spin detection (Dan Rugar, Talk on the 2004 IEEE NTC Quantum Device Technology Workshop). First we describe the quantum dynamics of the cantilever-spin system using simple estimates in the spirit of the mean field approximation. Then we present the results of our computer simulations of the Schrödinger equation for the wave function of the cantilever-spin system and of the master equation for the density matrix of the system. We demonstrate that the cantilever behaves like a quasi-classical measurement device which detects the spin projection along the effective magnetic field. We show that the OSCAR technique provides continuous monitoring of the single spin, which could be used to detect the mysterious quantum collapses of the wave function of the cantilever-spin system.

*Index Terms*—— Magnetic Resonance Force Microscopy (MRFM), adiabatic reversals, wave function collapse, quantum jumps, master equation, magnetic noise, micromechanical cantilever, quantum decoherence, thermal diffusion, quantum entanglement.

## I. INTRODUCTION

THE theory of single spin magnetic resonance force microscopy (MRFM) originated from John Sidles who proposed a way to measure the magnetic force produced by a single spin combining magnetic resonance, atomic force microscopy, and micromechanical resonance of the ultra-sensitive cantilever [1]. The practical implementation of this proposal would allow an atomic scale magnetic imaging below the surface of a nontransparent material. Optical as well as scanning tunneling microscopy detection of a single spin is restricted to the surface atoms. (See, for example, [2]-[3]). In his pioneering work John Sidles discussed the detection of a single nuclear spin. In reality even detection of a single electronic spin is a major challenge for the experimentalists: it requires measurement of a force of the order of a few attonewtons. The implementation of a single spin MRFM remained elusive until Dan Rugar and his team invented the oscillating cantilever-driven adiabatic reversals (OSCAR) technique and then demonstrated two spin sensitivity [4].

In this paper, we present the theory of the single spin OSCAR MRFM. In the second section, we describe the OSCAR dynamics in the spirit of the mean field approximation and estimate its characteristic parameters. We discuss the frequency shift of the cantilever vibrations, the thermal noise of the cantilever, the magnetic noise experienced by the spin, the opportunity of formation of the Schrödinger cat state, the decoherence, the quantum collapses of the wave function and the quantum jumps of the cantilever spin system. We also discuss the exciting possibility of measuring the characteristic time of the wave function collapse. In the third section, we present the results of our computer simulations of the spin-cantilever dynamics based on the Schrödinger equation for the wave function and the master equation for the density matrix of the spin-cantilever system. Recently other theoretical aspects related to the single-spin MRFM have been extensively discussed. (See, for example, ref. [5]-[13]).

## II. SPIN-CANTILEVER DYNAMICS IN OSCAR: DESCRIPTION AND ESTIMATIONS

### A. The Basic principles of the OSCAR technique

The main idea of the OSCAR MRFM technique invented by Dan Rugar and his team will now be summarized. An ultra-sensitive micromechanical cantilever (about 100 *nm* thick) with a ferromagnetic particle (about 1 $\mu m$ size) attached to the cantilever tip (CT) oscillates near the surface of a sample with a fixed amplitude. Fig.1 shows the OSCAR MRFM setup for the "perpendicular geometry." In its equilibrium position the cantilever is perpendicular to the sample surface.

When the CT with the ferromagnetic particle moves from the right endpoint of its trajectory to the left endpoint, the dipole field produced by the ferromagnetic particle on the spin decreases. Let us consider the effective magnetic field $\vec{B}_{ef}$ in the system of coordinates rotating with the *rf* field. The direction of the effective magnetic field reverses in the *x-z* plane from the *+z* to the *−z* direction. (We assume that the resonant condition $\omega = \gamma B$, where $\omega$ is the *rf* frequency, and $\vec{B}$ is the magnetic field on the spin, is satisfied for the

Manuscript received June 9, 2004. This work was supported by NSA and DARPA.
G.P. Berman is with the Los Alamos National Laboratory, NM 87544 USA (phone: (505)667-2489; fax: (505)665-3003; e-mail: gpb@ lanl.gov).
F. Borgonovi is with the University Cattolica, Brescia, Italy (e-mail: borgonov@dmf.bs.unicatt.it).
V.N. Gorshkov is with the Los Alamos National Laboratory, NM 87544 USA (e-mail: gorshkov@ cnls.lanl.gov).
V.I. Tsifrinovich is with Polytechnic University, New York (e-mail: vtsifrin@duke.poly.edu).

equilibrium position of the CT, and the rotating *rf* field points in the positive *x*-direction). If the condition of the adiabatic motion $|d\vec{B}_{ef}/dt| \ll \gamma B_1^2$ is satisfied the "average spin" $\langle \vec{S} \rangle$ follows the direction of the effective field.

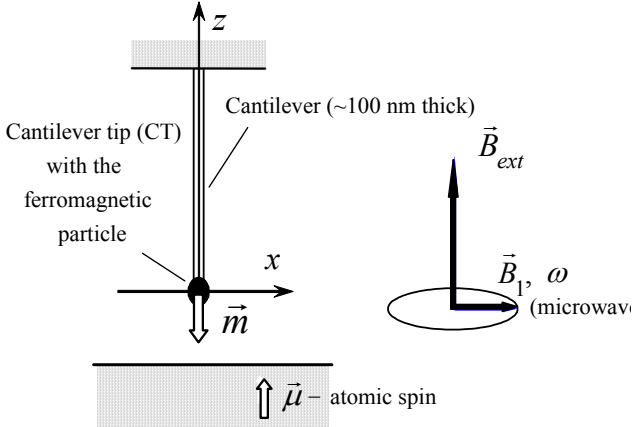

Fig. 1. The OSCAR MRFM setup for the perpendicular geometry. A ferromagnetic particle with the magnetic moment $\vec{m}$ is attached to CT and oscillates near the surface of a sample. $\vec{B}_{ext}$ is the external permanent magnetic field; $\vec{B}_1$ is the *rf* rotating field of frequency $\omega$. An atomic spin with the magnetic moment $\vec{\mu}$ is placed not far from the sample surface.

We assume the electronic spin is initially in its ground state, *i.e.* it points in the negative *z*-direction (the electronic spin points in the direction opposite to its magnetic moment). If the *rf* field is turned on when the CT is at its right end position, the effective field points initially in the positive *z*-direction. In the process of adiabatic motion, the spin remains anti-parallel to the effective field. The *z*-component of the spin magnetic moment $\vec{\mu} = -\gamma \langle \vec{S} \rangle$ oscillates with the CT frequency. It produces a back resonant magnetic force on CT: $F_x = G\mu_z$, where $G = |\partial B_z/\partial x|$ is the gradient of the magnetic field at the spin (here and below when we mention CT we mean the CT including the ferromagnetic particle; certainly, the magnetic force is acting on the ferromagnetic particle). Since $\mu_z$ is proportional to the CT displacement from the equilibrium, the magnetic force is also proportional to this CT displacement. Thus, the magnetic force influences the effective spring constant of CT and consequently the CT frequency (which is the fundamental frequency of the cantilever). The CT frequency shift can be measured with high accuracy – this is the main advantage of the OSCAR technique. The direction of the magnetic force acting on CT is opposite to the direction of the spring force. Thus, for the electron spin pointing opposite to the effective field, the CT frequency will decrease. If the electron spin points in the direction of the effective field, the CT frequency will increase.

Under the conditions of adiabatic motion, the spin component along the effective field is an approximate integral of motion. Thus, we may consider the CT as a quasi-classical device which measures this spin component. However there is a very important point: the CT continuously monitors the state of the spin. Thus, we may expect the outcome of the OSCAR MRFM shown in Fig. 2: the CT frequency shift takes one of the two values $|\delta\omega_c|$ or $-|\delta\omega_c|$, depending on the direction of the spin relative to the effective field. Quantum jumps of the spin cause jumps in the CT frequency shift.

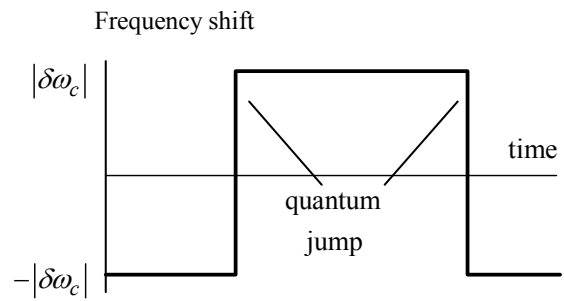

Fig. 2. Expected output of a single spin OSCAR MRFM. $\delta\omega_c$ is the frequency shift of the CT vibrations.

To increase the measurement sensitivity, Dan Rugar and his team implemented a modified technique which is called the "interrupted OSCAR" technique. They interrupted the *rf* field periodically (with a period $T_i$ of about 10 *ms*). When the CT was at its end point, the applied *rf* field was interrupted for a time interval equal to half of the CT vibration period. At the end of the "dead interval" the effective field reverses while the spin retains its initial direction. This effect is equivalent to the application of the effective $\pi$-pulse in the rotating frame. As a result the CT frequency shift becomes a periodic function of time with twice the interruption period $2T_i$. Now the OSCAR signal is detected at the frequency $1/(2T_i)$.

### B. Estimation of the Frequency

For our estimations we will use the values of parameters from ref. [4]. (Although the experiment in [4] was conducted with many spins, its setup with two-spin sensitivity is probably appropriate for single-spin detection):

The effective CT spring constant $k_c = 600\,\mu N/m$,

The CT frequency and period $f_c = \omega_c/2\pi = 6.6\,kHz$,

$T_c = 150\,\mu s$,

The CT quality factor $Q = 5\times10^4$,

The CT amplitude $A = 10\,nm$,

The rotating rf field amplitude and frequency $B_1 = 300\,\mu T$, $\omega = 3\,GHz$, $(\omega/\gamma = 100\,mT)$,

The Rabi frequency and period
$f_R = \omega_R/2\pi = \gamma B_1/2\pi = 8.4\,MHz$, $T_R = 120\,ns$,

The magnetic field gradient at a spin location
$G = |\partial B_z/\partial x| = 430\,kT/m$,

The maximum magnetic force on CT
$(F_x)_{max} = G\mu_B = 4\,aN$,

Temperature $T = 200\,mK$,

The correlation time for the CT frequency shift $\tau_m = 3\,s$.

We now estimate the CT frequency shift in the spirit of the mean field approximation. Let the spin be anti-parallel to the effective magnetic field $\vec{B}_{ef}$. Then

$$\langle S_z \rangle / S = -(B_{ef})_z / B_{ef}, \quad (1)$$

where $\vec{B}_{ef} = \{B_1, 0, G\langle x \rangle\}$. The net force on CT is given by

$$F_x = -k_c \langle x \rangle - \gamma \hbar G \langle S_z \rangle. \quad (2)$$

Combining these formulas and averaging over fast oscillations ($\langle x \rangle^2 \to A^2/2$) we obtain expressions for the relative shift of the effective spring constant and the frequency shift,

$$\delta k_c = -\gamma \hbar G^2 / [2(G^2 A^2 + B_1^2)]^{1/2},$$
$$\delta f_c / f_c = \delta k_c / (2k_c), \quad (3)$$

which correspond to the numerical values $\delta f_c/f_c = -4.7 \times 10^{-7}$ and $\delta f_c = 3\,mHz$. For our values of parameters $GA \gg B_1$, and the expression for $\delta k_c$ can be simplified:

$$\delta k_c = -\sqrt{2} G \mu_B / A. \quad (4)$$

This expression has a clear physical meaning: the magnetic force on CT cannot be greater than $G\mu_B$. That is why the shift of the CT spring constant and the frequency shift increases with decrease of the amplitude $A$.

Now we discuss the possibility of reducing the CT amplitude and increasing the CT frequency shift. The condition for the full adiabatic reversals can be represented as follows

$$1 \ll GA/B_1 \ll f_R/f_c. \quad (5)$$

The left inequality is the condition for full spin reversals (between $+z$ and $-z$ directions). The right inequality is the condition for adiabatic spin motion. For our parameters $GA/B_1 = 14$, and $f_R/f_c = 1270$. To increase the CT frequency shift we may sacrifice the full spin reversals retaining the adiabatic motion. Fig. 3 shows the partial reversals of the effective field.

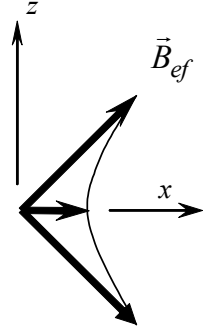

Fig. 3. Partial reversals of the effective field.

The use of partial adiabatic reversals is convenient for computer simulations because it allows us to save computational time. Below we show that this idea is not appropriate for the experiment as the thermal frequency noise also increases with decreasing CT amplitude.

III. THE INTERACTION BETWEEN THE CT-SPIN SYSTEM AND ITS ENVIRONMENT.

While the spin is parallel or anti-parallel to the effective field, the main manifestation of the CT-environment interaction in OSCAR is the thermal frequency noise. Now we will estimate its value. The *rms* coordinate of CT and the corresponding *rms* force are given by

$$x_{rms} = (k_B T/k_c)^{1/2},$$
$$F_{rms} = 2 k_c x_{rms}/Q. \quad (6)$$

To estimate the characteristic thermal spring constant fluctuation $\delta k_c^T$ we assume that the "thermal force" increases from 0 to $F_{rms}$ when the CT coordinate $x$ changes from 0 to $A$. Thus, $\delta k_c^T = F_{rms}/A$, and correspondingly, the characteristic thermal frequency fluctuation becomes

$$\delta f_c^T / f_c = x_{rms}/AQ. \quad (7)$$

The corresponding numerical values are $x_{rms} = 68\,pm$, $F_{rms} = 1.6\,aN$, and $\delta f_c^T/f_c = 1.4 \times 10^{-7}$. The estimated characteristic CT thermal frequency fluctuation is smaller than the OSCAR shift $\delta f_c$. On the other hand, one can see that thermal frequency fluctuation like the OSCAR frequency shift increases with the decreasing CT amplitude. Thus, the partial adiabatic reversals will not increase the signal-to-noise ratio.

Next, we consider the effect of the spin-environment interaction. This interaction can be described in terms of magnetic noise acting on the spin. Roughly speaking this noise causes a deviation of the spin from the effective field. This deviation generates two CT trajectories corresponding to the two possible direction of the spin relative to the effective field. These two trajectories manifest the formation of the Schrödinger cat state. Now the CT-environment interaction comes into the play. CT-environment entanglement quickly destroys the Schrödinger cat state leaving only one of the two possible trajectories. Physically this appears as a quantum collapse. Usually, the collapse pushes the spin back to the "pre-collapse" direction relative to the effective field. Sometimes the spin changes its direction. When a change occurs, we can observe the quantum jump by measuring the sharp change of the CT frequency shift.

Currently, the time of collapse is not predictable. We believe that understanding the timing of quantum collapses is one of the most interesting remaining problems in quantum theory. Let us assume that the collapse occurs when the separation between the two CT trajectories is of the order of the quantum uncertainty of the CT position $X_q$ (by 'CT position' we mean the position of the center-of-mass of the ferromagnetic particle)

$$X_q = (\hbar \omega_c / k_c)^{1/2}. \quad (8)$$

In this case the characteristic collapse time $t_{col}$ is of the order of the CT period $T_c = 180 \mu s$. (If we assume that the collapse occurs when the separation between the two trajectories is about $x_{mrs}$, which seems very unlikely, then $t_{col} \sim 10^4 T_c$.) The CT decoherence time $t_d$ can be estimated as

$$t_d = \omega_c \hbar^2 Q / (k_c k_B T \Delta x^2), \quad (9)$$

where $\Delta x$ is the separation between the two trajectories. (See, for example, [5]). Taking $\Delta x$ to be equal to the quantum uncertainty $X_q$ we obtain $t_d = 2 \mu s$. Thus, we have a typical quasi-classical systems situation: the decoherence time is much smaller than the time of separation of two trajectories, which is the time of formation of the Schrödinger cat state. (This is why the Schrödinger cat state is so elusive for quasi-classical systems.) It indicates that the collapse time depends on the CT frequency shift rather than on the decoherence time.

Next, we will estimate the characteristic time interval between two quantum jumps, $t_{jump}$. We assume that the most important source of the magnetic noise for the spin is associated with the cantilever modes whose frequencies are close to the Rabi frequency of the spin. (See, for example, [7], [8], [12]). The reason is the following. When the spin changes its direction between +z and −z, its frequency in the rotating frame changes between its maximum value $\omega_{\max}$ and its minimum value, which is the Rabi frequency, $\omega_R$. Because all cantilever modes have the same thermal energy $k_B T / 2$ the thermal amplitude of the mode is inversely proportional to its frequency. Thus, the greatest amplitude of the CT thermal vibrations is associated with the modes near the Rabi frequency. As an estimate, we consider those modes in the interval between the Rabi and twice the Rabi frequency. The CT thermal amplitude of the Rabi frequency is

$$A_R^T = (f_c / f_R)(2 k_B T / k_c)^{1/2} = 75 fm. \quad (10)$$

We estimate the correlation time to be the Rabi period $T_R$ and find the following characteristic angular deviation during the correlation time:

$$\Delta \theta_0 = \gamma T_R G A_R^T = 6.8 \times 10^{-4} rad. \quad (11)$$

The time of passing the frequency interval $(f_R, 2 f_R)$ is

$$\Delta t_1 = 3.4 f_R / (\gamma G A f_c) = 5.8 \mu s. \quad (12)$$

Assuming a diffusion process, we can estimate the square of the angular deviation during a single reversal: $\langle \Delta \theta_1^2 \rangle = D \Delta t_1$, where $D = \Delta \theta_0^2 / T_R$ is the diffusion coefficient. The angular deviation between the two collapses is

$$\langle \Delta \theta_{col}^2 \rangle = \langle \Delta \theta_1^2 \rangle t_{col} / (T_c / 2), \quad (13)$$

where $t_{col}$ is the characteristic time between the two collapses. The probability of a quantum jump is approximately

$$P_{jump} = \langle \Delta \theta_{col}^2 \rangle / 4. \quad (14)$$

Using the estimate $P_{jump}(t_{jump} / t_{col}) = 1$, we obtain

$$t_{jump} = t_{col} / P_{jump} = A / \left[ 1.7 \gamma G (A_R^T)^2 \right] = 14 s. \quad (15)$$

The experimental value of the frequency shift correlation time in [4] was found to be 3s but it was obtained for a group of spins, not for a single spin.

Note that the collapse time cancels out in the final expression for the jump time. One of the most mysterious phenomena of the quantum physics – the wave function collapse – remains elusive. Let us now discuss the possibility of measuring the collapse time in the OSCAR dynamics. Between two collapses the second CT trajectory appears inside the quantum uncertainty of the first trajectory. Because the two trajectories have the opposite sign of the frequency shift, the overall CT frequency shift is expected to decrease (in absolute value). More rigorously, the shift, $\delta t_j$, of the time interval between two consecutive passings of the equilibrium CT position is expected to decrease in absolute value:

$$|\delta t_j| < \delta t_c = \pi \delta \omega_c / \omega_c^2 . \qquad (16)$$

This change, in principle, could be measured experimentally. However our estimate shows that the expected change is very small. Defining the probability of two trajectories before the collapse as $P_1$ and $P_2$ we obtain the estimates:

$$\delta t_j = \delta t_c (P_1 - P_2),$$
$$|\delta t_j - \delta t_c| / \delta t_c = \langle \Delta \theta_{col}^2 \rangle / 2 = 2 \times 10^{-5} \qquad (17)$$

for $t_{col} = T_c$. The effect is negligible because the probability of the second trajectory occurring inside the quantum uncertainty of the first trajectory is very small. To resolve this obstacle we propose using a modification of the interrupted OSCAR. Assume one interrupts the microwave for a time interval equal to one quarter of the period of the CT vibrations. This interruption acts like an effective $\pi/2$-pulse in the rotating frame: it generates an angle of $\pi/2$ between the effective field and the spin. Now before the collapse, the probabilities of both trajectories inside the common quantum uncertainty are the same, and the frequency shift (more rigorously the shift of $\delta t_j$) is equal to zero. This large change in the frequency shift could probably be detected experimentally.

## IV. SIMULATIONS OF THE OSCAR MRFM

### A. Schrödinger Dynamics

We start from the Schrödinger description of the CT-spin system. The Hamiltonian of the system in the rotating reference frame is:

$$H = (p_x^2 + x^2)/2 + \varepsilon S_x + 2\eta x S_z + \Delta(t) S_z ,$$
$$\varepsilon = f_R / f_c , \quad \eta = \gamma G X_q /(2\omega_c) . \qquad (18)$$

The first term describes the CT motion, the second term is the interaction between the spin and the rf field, the third term is the CT-spin interaction, and the last term describes the effects of magnetic noise on the spin due to the spin-environment interaction. As we have mentioned before, the most important source of the magnetic noise is normally associated with the cantilever modes near the Rabi frequency. This magnetic noise causes a deviation of the spin from the effective field primarily when the spin passes through the transversal plane. That is why we consider only the $z$-component of the magnetic noise field. We do not include the CT-environment interaction because the main effect of this interaction – the decoherence – cannot be described in the scope of the Schrödinger equation.

In our simulations we use the following units:

Frequency: $f_c = 6.6 kHz$,
Length: $X_q = 85 fm$,
Momentum: $\hbar / X_q = 1.2 \times 10^{-21} Ns$,
Time: $1/\omega_c = 24 \mu s$,
Temperature: $\hbar \omega_c / k_B = 320 nK$.

The experimental values of our parameters in these units are the following: the amplitude $A = 1.2 \times 10^5$, the temperature $T = 6.25 \times 10^5$, $\varepsilon = 1270$, $\eta = 0.078$. Unfortunately these values of parameters are outside the scope of our computer capabilities. Thus, we simulate the CT-spin dynamics taking $A = 13$, $\varepsilon = 10$, $\eta = 0.3$. The conditions for the full adiabatic reversals in terms of our parameters,

$$\varepsilon \ll 2\eta A \ll \varepsilon^2 , \qquad (19)$$

are clearly violated because $2\eta A = 7.8$. In this case, we have partial adiabatic reversals with a relatively large CT frequency shift

$$|\delta f_c| = \eta^2 /(2\eta^2 A^2 + \varepsilon^2)^{1/2} = 8 \times 10^{-3} . \qquad (20)$$

The wave function of the system is a spinor $u_s(x,t)$, where the spin variable $s$ takes the two values $s = \pm 1/2$. (We use the $S_z$-representation). The initial wave function was taken in the form of the product of the CT part, $u_c(x)$, and the spin part, which describes the direction of the average spin. The CT part of the wave function describes the quasi-classical coherent state,

$$u_c(x) = \sum A_n u_n(x), \ A_n = \left[ \alpha^n /(n!)^{1/2} \right] \exp(-|\alpha|^2 /2),$$
$$\alpha = (\langle x(0) \rangle + i \langle p_x(0) \rangle)/\sqrt{2}, \ \langle x(0) \rangle = A, \ \langle p_x(0) \rangle = 0, \qquad (21)$$

where $u_n(x)$ are the eigenfunctions of the harmonic oscillator Hamiltonian.

First, we consider the CT-spin dynamics with no magnetic noise. If the initial average spin points opposite to the effective field $\{\varepsilon, 0, 2\eta\langle x(t)\rangle\}$, then the wave function remains to be the product of the CT and spin parts. The probability density $P = \sum_s |u_s(x,t)|^2$ represents a single peak oscillating with the frequency, $(1-|\delta\omega_c|)$. The value of $|\delta\omega_c| = 7.9 \times 10^{-3}$ is very close to the estimated value. The same is true for the initial average spin pointing in the direction of the effective field. The only difference is the frequency of oscillations, which is now equal to $(1+|\delta\omega_c|)$. If the initial average spin makes an angle $\pi - \theta$ with the effective field (see Fig. 4) the wave function describes an entangled CT-spin state.

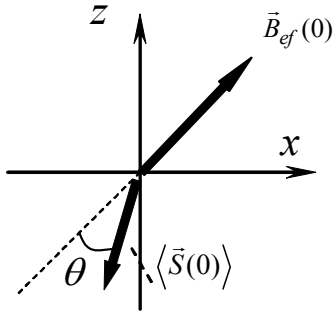

Fig.4. The initial average spin makes an angle $\pi - \theta$ with the effective field.

The probability density peak gradually splits into two peaks. (See Fig. 5). Thus, the wave function describes the Schrödinger cat state of the CT-spin system.

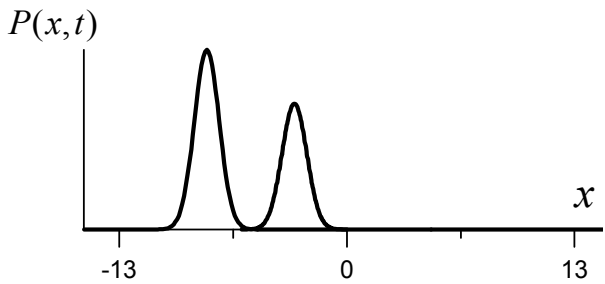

Fig. 5. The Schrödinger cat state of the CT-spin system.

The first peak oscillates with the frequency $(1-|\delta\omega_c|)$, and the second peak oscillates with the frequency $(1+|\delta\omega_c|)$. Note that both components of the spinor $u_s(x,t)$ contribute to every peak. If we consider only the part of the wave function describing one of the peaks then it can be decomposed into the product of the CT part and the spin part with the average spin pointing in (or opposite to) the direction of the effective field corresponding to this peak. (Note that the effective fields corresponding to the two peaks are not anti-parallel to each other.) The area under the first peak is $\sin^2(\theta/2)$, and the area under the second is $\cos^2(\theta/2)$. All these facts prove that the CT-spin dynamics exhibits the Stern-Gerlach effect: the two spin directions relative to the effective field generate two separate CT trajectories.

Next we include the effect of the magnetic noise. We assume the noise field, $\Delta(t)$, to be a random telegraph signal with amplitude $\Delta_0$. The time interval between two "kicks" of the noisy field is taken randomly from the interval, $(T_R - T_R/4, T_R + T_R/4)$. The initial average spin points opposite to the effective field. Fig. 6 demonstrates the shift of the time interval between two consecutive passes through the CT equilibrium position before the split of the two CT trajectories. One can see a decrease in the time interval shift for $\Delta_0 = 0.3$ and $\Delta_0 = 0.5$. (For the experimental value, $\Delta_0 = GA_R^T = 0.13$, this effect is negligible.)

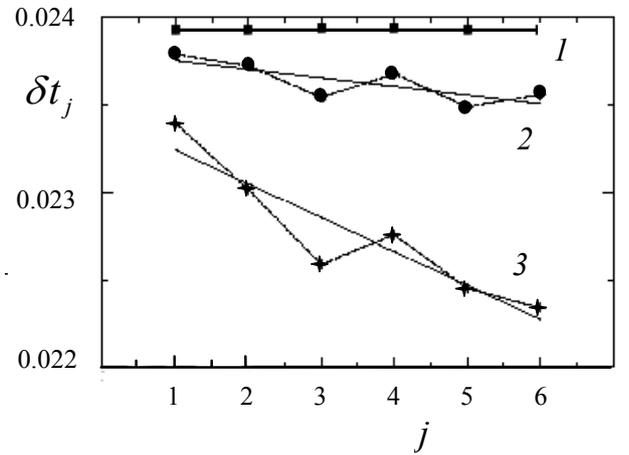

Fig. 6. The shift of the time interval between the consecutive passes through the CT equilibrium positions for $\Delta_0 = 0$ (1), $\Delta_0 = 0.3$ (2), and $\Delta_0 = 0.5$ (3). The estimated value of the shift is $\delta t_c = \pi\delta\omega_c = 0.0248$.

Gradually the probability peak splits into two peaks but in this case the two trajectories are generated by the noisy field rather than the initial conditions.

### B. Master Equation

In order to describe the CT decoherence and the thermal diffusion we consider an ensemble of the CT-spin systems interacting with the environment. We use the Caldeira-Leggett master equation for the density matrix:

$$\frac{\partial}{\partial \tau} \rho(x, x', s, s', t) = \hat{L} \rho(x, x', s, s', t) - i\frac{\varepsilon}{2}[\rho(x, x', s, -s', t) - \rho(x, x', -s, s', t)],$$

$$\hat{L} = \frac{i}{2}\left(\frac{\partial^2}{\partial x^2} - \frac{\partial^2}{\partial x'^2}\right) - \frac{i}{2}\left(x^2 - x'^2\right) - \frac{1}{2Q}(x - x') - \frac{T}{Q}(x - x')^2 - 2i\eta(x's' - xs).$$

(22)

$$\rho(x, x', s, s', \tau) = \begin{pmatrix} R_{\frac{1}{2},\frac{1}{2}} & R_{\frac{1}{2},-\frac{1}{2}} \\ R_{-\frac{1}{2},\frac{1}{2}} & R_{-\frac{1}{2},-\frac{1}{2}} \end{pmatrix}$$

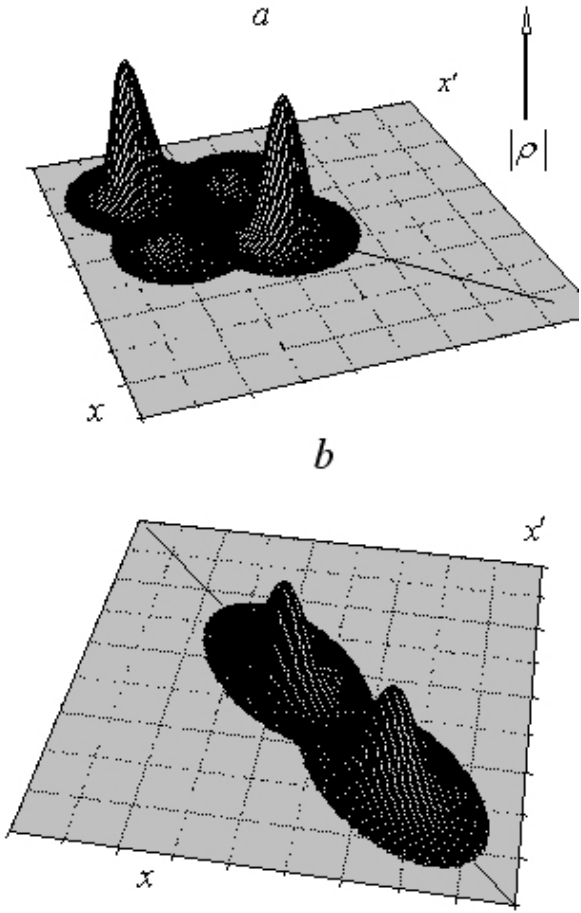

Fig. 7. The typical shapes for the modulus of the reduced CT density matrix (*a*) before - and (*b*) after the disappearance of the non-diagonal peaks.

We cannot demonstrate the CT decoherence if the decoherence time is smaller than the time of formation of two CT trajectories (the Schrödinger cat state). Thus, to simulate the decoherence, we use the very low temperature $T = 20$ (instead of the experimental value $T = 6.25 \times 10^5$). To save the computer time we use $A = 8$ and $Q = 1000$. The values of $\varepsilon$ and $\eta$ are the same those we have used for the Schrödinger dynamics, and $\Delta_0 = 0$. The initial density matrix is taken as a product of the CT and spin parts:

$$\rho(x, x', s, s', 0) = R(x, x')\Lambda(s, s'),$$
$$R(x, x') = u_c(x) u_c^*(x'), \ \langle x(0) \rangle = A, \ \langle p_x(0) \rangle = 0.$$
(23)

If the initial direction of the average spin is anti-parallel or parallel to the effective field, the density matrix remains a product of the CT part and the spin part. The modulus of the CT part describes a single peak, which oscillates along the diagonal $x = x'$ with the frequency $(1 - |\delta\omega_c|)$ or $(1 + |\delta\omega_c|)$. This peak spreads along the diagonal demonstrating the thermal diffusion of the ensemble. The situation changes if the initial average spin makes an angle $\theta$ with the effective field, $\theta \neq 0, \pi$. In this case the CT and the spin become entangled and the initial peak of the modulus of the reduced CT density matrix $|\rho| = \left|\sum_s \rho(x, x', s, s, t)\right|$ splits into four peaks. As an example, Fig. 7 demonstrates two typical shapes of the reduced CT density matrix if $\theta = \pi/2$.

After the split of the initial peak, one can observe non-diagonal peaks, which describe the coherence between the two CT trajectories (the Schrödinger cat state). These peaks quickly disappear demonstrating the CT decoherence. Subsequently, the remaining diagonal peaks describe the statistical mixture of the two CT trajectories corresponding to the two spin directions relative to the effective field. These peaks spread along the diagonal $x = x'$ demonstrating the thermal diffusion in the ensemble. If we consider only the part of the density matrix describing one of the diagonal peaks then it can be decomposed into the product of the CT and spin parts with the average spin pointing in (or opposite) the direction of the effective field corresponding to this peak. Thus the master equation allows us to simulate the CT-spin decoherence if the decoherence time is greater than the time of the Schrödinger cat formation. Also using the master equation we can demonstrate the process of the thermal diffusion.

## V. CONCLUSION

In this paper we have presented a theory of the single-spin OSCAR MRFM. We presented estimates for the three main experimental parameters of the OSCAR technique: the CT frequency shift, the frequency noise, and the characteristic time between the spin quantum jumps. We proposed an experiment for measuring the most elusive parameter: the average time interval between the collapses of the CT-spin wave function. We also reported the results of computer simulations using both the Schrödinger equation and the master equation. Our simulations demonstrate that CT can be considered as a quasi-classical device, which measures the spin direction relative to the effective magnetic field.

At the 2004 IEEE NTC Quantum Device Technology Workshop Dan Rugar reported the first experimental detection of a single atomic spin using OSCAR MRFM. This historical event marks the beginning of experimental single-spin imaging in condensed matter. The next step will be the continuous single-spin measurement. We hope that besides tremendous imaging and quantum information processing applications the OSCAR MRFM will allow one to measure one of the most mysterious events in the quantum physics – the collapse of the wave function.

ACKNOWLEDGMENT

The authors would like to thank Dan Rugar for valuable discussions.

**Gennady Berman** received his MS degree from the Novosibirsk State University, Russia, in 1970; a Ph..D. degree from the Kirensky Institute of Physics, in 1974, Krasnoyarsk, Russia; and a Second Doctor Degree in Physics and Mathematics, in 1989, from the Institute of General Physics, Moscow, Russia. He is currently Deputy Group Leader of the Complex Systems Group, Theoretical Division, LANL. He is co-author of five books and more than 150 papers on quantum chaos, nano-technology, quantum computation, and quantum measurement.

**Fausto Borgonovi** received his MS degree from the University of Milan, Italy in 1984, and his Ph.D. degree from the University of Pavia, Italy in 1989. He was a Postdoctoral Fellow at the University of Pavia in 1990-91, and researcher at C.N.R.S. , Laboratoire de Physique Quantique, UPS, Toulouse, France in 1993. He is an Assistant Professor (permanent position) since 1993 in "Theoretical Physics" at the Faculty of Science of Universita' Cattolica, Brescia, Italy. His fields of activity include Theoretical Physics, Quantum Chaos, Nonlinear Systems, Quantum Computation, Quantum Measurement and Decoherence.

**Vyacheslav Gorshkov** received his MS degree from the Rostov-on-Don State University, USSR, in 1970; his PhD degree from the Institute of Physics of National Academy of Sciences of Ukraine, Kiev, in 1976, and his Doctor of Sciences degree from the Institute of Radio-Physics and Electronics, Kharkov, in 1991. He was a Leading Researcher at the Department of the Gas-Electronics at the Institute of Physics and the Head of the Department of Applied Physics of the East-Ukrainian National University. He is now a Visiting Scientist at the Los Alamos National Laboratory. His current scientific interests include computer simulations of physical processes in electron-hole plasmas in semiconductors and plasmas of gas discharge; nonlinear electro-hydro-dynamics; plasma-optics; physics of singularities of optical beams; colloidal physics; and quantum dynamics.

**Vladimir Tsifrinovich** received his MS degree from the Krasnoyarsk State University, USSR, in 1972. He received his Ph.D and Doctor of Sciences degrees from the Institute of Physics of the Russian Academy of Sciences, Krasnoyarsk, Russia, in 1977 and 1992, respectively. He worked as a Research Scientist in the Institute of Physics of the Russian Academy of Sciences, before he joined Polytechnic University, New York, in 1994, where he is now Lecturer of Physics. He is also a co-investigator in two research projects in the Los Alamos National Laboratory. His current research interests include quantum computation, quantum measurements, and magnetic resonance. He is an editor of the *International Journal of Quantum Information.*